\documentclass[usenatbib]{mn2e}
\usepackage{graphicx, amssymb, aas_macros, natbib}
\newcommand{\mvir}{M_{\rm{vir}}}

\newcommand{\mstar}{{\rm M}_{\star}}
\newcommand{\lstar}{L^{*}}
\newcommand{\msun}{{\rm M}_{\odot}}

\newcommand{\mpc}{{\rm Mpc}}
\newcommand{\kpc}{{\rm kpc}}
\newcommand{\kms}{{\rm km \, s}^{-1}}

\title[The Mass Dependence of Satellite Quenching]
{
The Mass Dependence of Satellite Quenching in Milky Way-like Halos
}
\author[Phillips et al.]
{John I. Phillips,$^1$\thanks{$\!\!$e-mail: johnip@uci.edu} 
Coral Wheeler,$^1$ 
Michael C. Cooper,$^1$ 
Michael Boylan-Kolchin,$^2$
\newauthor James S. Bullock,$^1$ 
Erik Tollerud$^3$\thanks{$\!\!$Hubble Fellow}\\
$\!\!^1$Center for Cosmology, Department of Physics and Astronomy, 
  4129 Reines Hall, University of California, Irvine, CA 92697, USA\\
$\!\!^2$Astronomy Department, University of Maryland
College Park, MD 20742-2421\\
$\!\!^3$Astronomy Department, Yale University, P.O. Box 208101, New Haven, CT
06510, USA}
\begin{document}

\pagerange{\pageref{firstpage}--\pageref{lastpage}} 
\pubyear{2014}

\maketitle

\label{firstpage}
\begin{abstract}
  Using the Sloan Digital Sky Survey, we examine the quenching of
  satellite galaxies around isolated Milky Way-like hosts in the local
  Universe. We find that the efficiency of satellite quenching around
  isolated galaxies is low and roughly constant over two orders of
  magnitude in satellite stellar mass ($\mstar$ = $10^{8.5}-10^{10.5}
  \, \msun$), with only $\sim~20\%$ of systems quenched as a result of
  environmental processes. While largely independent of satellite
  stellar mass, satellite quenching does exhibit clear dependence on
  the properties of the host. We show that satellites of passive hosts
  are substantially more likely to be quenched than those of
  star-forming hosts, and we present evidence that more massive halos
  quench their satellites more efficiently. These results extend
  trends seen previously in more massive host halos and for higher
  satellite masses. Taken together, it appears that galaxies with
  stellar masses larger than about $10^{8}~\msun$ are uniformly
  resistant to environmental quenching, with the relative harshness of
  the host environment likely serving as the primary driver of
  satellite quenching. At lower stellar masses ($< 10^{8}~\msun$),
  however, observations of the Local Group suggest that the vast
  majority of satellite galaxies are quenched, potentially pointing
  towards a characteristic satellite mass scale below which quenching
  efficiency increases dramatically. 
\end{abstract}


\section{Introduction}
\label{sec:Intro} 

It is well documented that galaxy properties, such as morphology and
star-formation rate, depend upon the local galaxy density, often
referred to as the ``environment'' in which a galaxy is located. For
instance, passive systems are systematically overrepresented in
high-density environments at both low and intermediate redshift
\citep[e.g.][]{davis76, dressler80, lewis02, balogh04, hogg04,
  cooper06, cooper07, cooper08, cooper10b, cooper10a}. This observed
dependence of galaxy properties on local environment is most apparent
at lower stellar masses, with satellite galaxies in groups and
clusters exhibiting redder rest-frame colors, more bulge-dominated
morphologies, as well as older and more metal-rich stellar populations
than their counterparts of equal stellar mass in the field
\citep[e.g.][]{baldry06, yang07, vdb08, peng10, pasquali10, woo13}.

A variety of physical mechanisms are potentially responsible for the
generally lower star-formation rates and higher incidence of bulges
for satellite galaxies relative to their field counterparts. In
particular, processes such as strangulation
\citep{larson80,balogh00a,balogh00b}, ram-pressure stripping
\citep{gunn72,quilis00}, and harassment \citep{farouki81,moore96} may
preferentially suppress star formation and transform the structure of
satellite galaxies in more massive halos (i.e.~higher-density
environments).  At present, our understanding of which mechanism(s)
dominate(s) the evolution of low-mass satellite galaxies is woefully
incomplete. For example, modern semi-analytic models of galaxy
evolution dramatically overpredict the number of passive satellite
galaxies in the local Universe \citep{kimm09, weinmann10, weinmann12}
and many hydrodynamical models fail to reproduce the low-mass end of
the stellar mass function at $z \sim 0$ \citep{crain09, dave11}.

Understanding the mass dependence of satellite quenching represents a
critical step towards identifying the specific physical processes at
play in driving the evolution of low-mass galaxies as a function of
environment. Simple models of ram-pressure stripping and strangulation
predict an increased efficiency at low mass, such that less-massive
satellites should be more readily quenched by these processes. In
addition, understanding the dependence of environmental effects on
host mass is similarly fundamental, as more massive host halos are
likely to harbor a hotter and denser circumgalactic medium that may
more efficiently strip or strangulate infalling satellite systems.
In a previous paper \citep[][hereafter P14]{P14}, we pointed out a
significant trend in satellite galaxy quenching that may result from
this dependence on host mass: passive Milky Way-mass galaxies quench
their satellites whereas star forming Milky Way-mass galaxies do
not. Specifically, for massive satellites (with $\mstar \sim \!
10^{10}~\msun$), \emph{passive} $\sim \! L^{*}$ hosts quench roughly
$30\%$ of their infalling satellites, while the satellites of
\emph{star-forming} $\sim \! L^{*}$ hosts exhibit the same
star-forming activity as a field sample with the same stellar mass
distribution. Using stacked line-of-sight satellite kinematics to
estimate the host halo mass for these systems suggests that this
dichotomy in quenching may be partially related to larger dark matter
halo masses for passive $L^{*}$ hosts relative to their star-forming
counterparts at $z \sim 0$.

In this work, we expand upon the analysis of P14 by examining trends
in satellite quenching beyond the dichotomy in host star-formation
activity, focusing on the mass dependence of satellite quenching for
Milky Way-like hosts. In particular, we compare satellites with
stellar masses of $10^{8.5}-10^{10.5}~\msun$ around a
carefully-selected sample of isolated $\lstar$ galaxies to field
galaxies with equivalent stellar mass and/or specific star-formation
rate (SSFR) that are the central galaxy in their respective dark
matter halo. By focusing on isolated hosts, we are able to probe
systems residing in dark matter halos comparable to that of our Milky
Way \citep[${\rm a~few}~\times~10^{12}~\msun$,][]{deason12, bk13,
  vdm12}, while eliminating the known effects of more massive halos
and large-scale structure on satellites. As motivated by the now
well-established bimodality of galaxies in color-versus-luminosity
space (i.e.~the red sequence and blue cloud) and the dichotomy of
quenching for which we argue in P14, we will often separately consider
trends where the central host galaxy is passive from those where the
central galaxy is star-forming. By characterizing the mass dependence
of satellite quenching, we aim to constrain the physical mechanisms
dominating the evolution of satellite systems in Milky Way-like halos.

The paper is structured as follows: In Section \ref{sec:sim}, we
describe our methodology for identifying Milky Way-like halos and
describe the specific criteria applied to create our central/host and
satellite samples as well as our control samples. In Section
\ref{sec:results}, we present our primary findings on the dependence
of satellite quenching efficiency on the properties of the satellite
and host galaxies. Finally, in Section \ref{sec:discuss}, we discuss
the implication of the observed trends on galaxy formation
models. Throughout our analysis, we employ a $\Lambda$ cold dark
matter ($\Lambda$CDM) cosmology with WMAP7+BAO+H0 parameters
$\Omega_{\Lambda} = 0.73$, $\Omega_{m} = 0.27$, and $h =0.70$
\citep{BAO}. Unless otherwise noted, all logarithms are base 10, and
all quoted virial masses are derived from a spherical top-hat model
according to the stellar mass/virial mass relation given in
\citet{guo11}.

\section{Sample Selection}
\label{sec:sim}

In selecting our observational sample, we employ data from Data
Release 7 (DR7) of the Sloan Digital Sky Survey
\citep[SDSS,][]{york00, abazajian09}. In particular, we utilize the
MPA-JHU derived data products, including median total stellar masses,
photometrically derived according to \citet[][see also
\citealt{salim07}]{kauffmann03}, and median total star formation
rates, measured from the SDSS spectra as detailed by
\citet{brinchmann04}. Spectroscopic completeness information as a
function of position on the sky (i.e.~\emph{fgotmain}) is drawn from
the NYU Value-Added Galaxy Catalog
\citep[NYU-VAGC,][]{blanton05vagc}. Our selection criteria follow very
closely those of P14, motivated by careful analysis of the Millennium
II simulation \citep[MS-II,][]{bk09}. We will summarize them here; for
a full discussion we refer the reader to that paper. Throughout this
work, we refer to objects associated on the sky as ``primaries'' and
``secondaries,'' while populations that have undergone correction for
the presence of interlopers will be referred to as ``hosts'' and
``satellites,'' as the corrected data more accurately reflect the
properties of true host-satellite systems.

\subsection{Sample Selection}

For our sample of primaries, we select all objects in the SDSS
spectroscopic sample with a stellar mass of $\mstar > 10^{10.5}~\msun$
and with $z < 0.032$, restricting to SDSS fiber plates with a
spectroscopic completeness of $> \! 0.7$ for the main galaxy
sample. We then apply the following isolation criteria to
preferentially select galaxies residing in halos of mass comparable to
that of the Milky Way (${\rm a~few}~\times~10^{12}~\msun$), where the
halo mass of host systems in our sample is calibrated by applying our
selection criteria to the MS-II simulation. First, we allow no other
galaxies with a stellar mass of $\mstar > 10^{10.5}~\msun$ within a
cylinder defined by a radius of $350$~kpc in projection and a length
in velocity space of $2000~\kms$ along the line-of-sight and centered
on the primary. In addition, we define an annulus with an inner
(outer) radius of $350~\kpc$ ($1~\mpc)$, wherein we allow no more than
one galaxy of stellar mass $\mstar > 10^{10.5}~\msun$. Galaxies that
pass these criteria are deemed isolated primaries.

To select spectroscopically-confirmed satellites or secondaries around
our isolated primaries, we define a search region with a radius of
$350~\kpc $ on the sky and $\pm~500~\kms $ in velocity space. We
search for secondaries in two mass ranges: ``massive secondaries" with
stellar mass of $10^{9.5}~\msun < \mstar < 10^{10.5}~\msun$ and a
maximum redshift of $z = 0.032$, and ``dwarf secondaries" with stellar
mass of $10^{8.5}~\msun < \mstar < 10^{9.5}~\msun$ and a maximum
redshift of $z = 0.024$. These limiting redshifts are designed to
ensure that we are complete at all stellar masses and SSFRs under
consideration.

\begin{figure}
 \centering
 \includegraphics[scale=0.36, viewport=20 0 800 410]{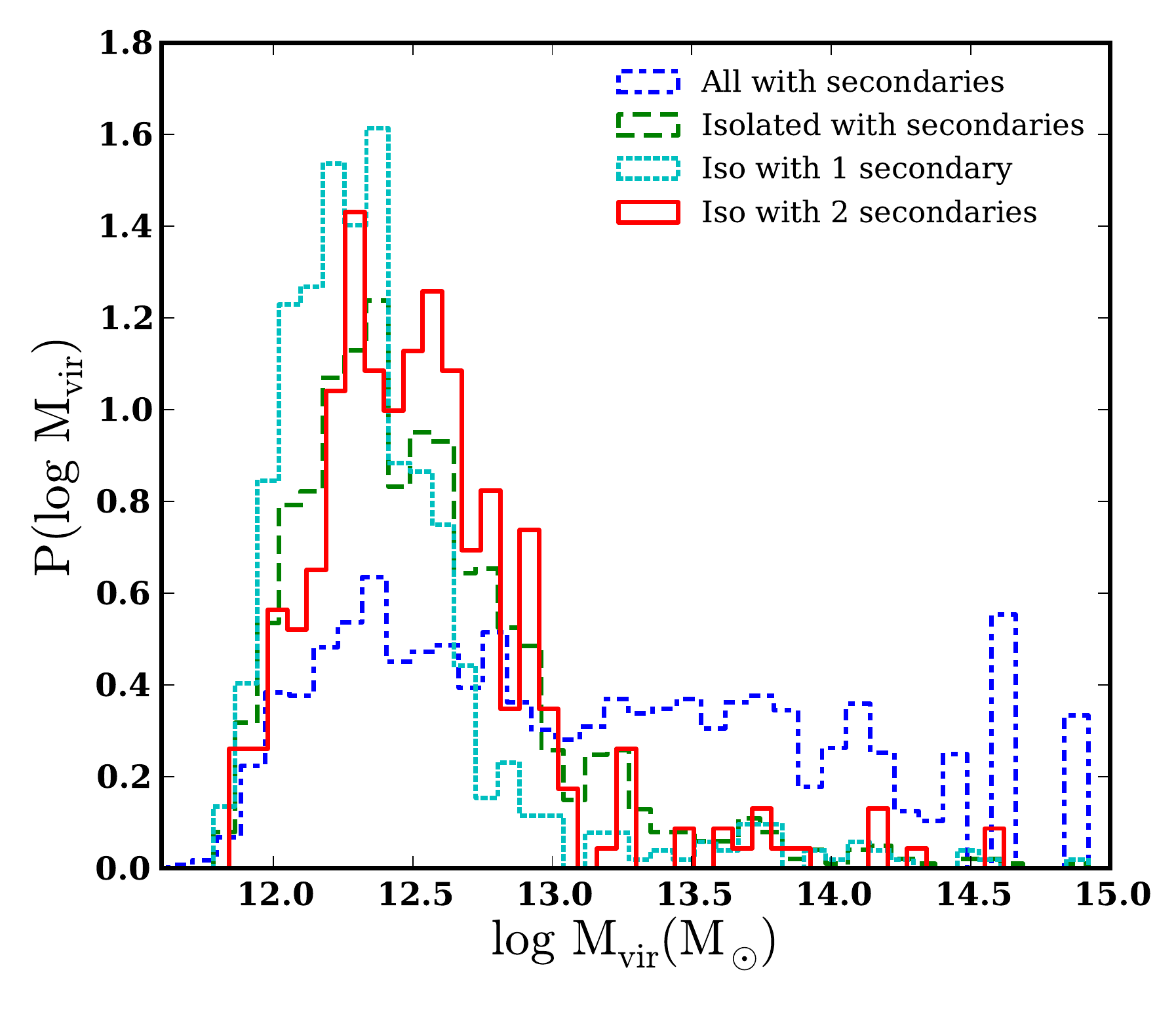}
 \caption{Virial mass distributions illustrating our host selection
   criteria. Shown are the virial mass distributions for all primaries
   with secondaries (blue short-dashed line), primaries that pass our
   isolation criteria and have at least one secondary (dashed green
   line), primaries that pass our isolation criteria and have exactly
   two secondaries (solid red line), and primaries that pass our
   isolation criteria and have exactly one secondary (dotted cyan
   line). Applying our selection criteria and restricting the number
   of massive satellites effectively removes massive halos from the
   sample.  Primaries with fewer massive satellites tend to have lower
   virial masses. As shown by P14, restricting the sample to primaries
   with only one massive secondary efficiently selects Milky Way-like
   systems.  }
 \label{fig:mtop}
\end{figure}

According to the number of massive secondaries identified about each
primary, we divide our sample of isolated primaries, such that our
main sample consists of primaries with exactly one secondary. As
discussed by P14, restricting to systems with exactly one massive
secondary in the SDSS identifies halos with a mass distribution
sharply peaked at ${\rm a~few}~\times~10^{12}~\msun$. We also consider
primaries with exactly zero and exactly two massive
secondaries. However, primaries with three or more massive secondaries
are excluded, since such systems are strongly biased towards the
group/cluster regime. Figure~\ref{fig:mtop} shows the virial mass
distributions for primaries in our mass range from comparison to the
MS-II simulation.\footnote{For further details regarding the analysis
  of the MS-II simulation, we refer the reader to the detailed
  discussion in P14 as well as Section \ref{sec:interlop} herein.}
Applying our isolation criteria removes most primaries residing in
clusters, while restricting the sample to primaries with at most two
massive satellites strongly selects against systems with
$\mvir~>~10^{13}~\msun$. Primaries with two massive satellites
represent a subsample of isolated objects with greater virial masses,
such that these systems are more likely to have
$\mvir~\gtrsim~10^{12.4}~\msun$ than primaries with a single massive
satellite. We give no consideration to the number of dwarf secondaries
orbiting a primary.

For both of the secondary samples, we construct corresponding control
samples of isolated field galaxies over the same mass and redshift
ranges, which we will refer to as the ``massive control" and ``dwarf
control" samples, respectively. These samples are subjected to an
isolation procedure more rigorous than that of the primaries; for both
samples, we require that no galaxy more massive than the lower mass
limit of the respective sample be within $3~\mpc$ on the sky and
$\pm~400~\kms$ in velocity. This ensures that the control samples are
almost completely comprised of objects that are themselves the primary
galaxy in their dark matter halo (${\rm f}_{\rm purity} \sim 97\%$,
see P14 for a full discussion of purity considerations) --- i.e.~the
samples are free of satellite galaxies. The number of objects in each
of our samples, including control samples, is given in Table~1.

\begin{table*}
\centering 
\begin{tabular}{l c c c} 
\hline\hline 
 Mass Range & Sample & N & $<\rm log \, M_{\star}>$ 
\\ [0.5ex] 
\hline 
 &One Satellite (``Main Sample")& 457 & 9.98 \\  \raisebox{1.5ex} {Massive}& Two Satellites & 306 & 9.98 \\ \raisebox{1.5ex}{$10^{9.5} M_{\odot} < M_{\star} < 10^{10.5} M_{\odot}$}   & Control & 581 & 9.94\\[1ex] 
\hline 
Dwarf &Satellite & 665 & 8.95\\ $10^{8.5} M_{\odot} < M_{\star} < 10^{9.5} M_{\odot}$  & Control & 302 & 9.16\\[-1ex] 
\hline

\label{tab:PPer} 
\end{tabular} 
\caption{Number of galaxies and mean log $\mstar$ in the satellite and control samples used
  in this study. Massive and dwarf satellites (along with
  their respective control samples) are restricted to stellar masses
  of  $10^{9.5}~\msun < \mstar < 10^{10.5}~\msun$ and $10^{8.5}~\msun
  < \mstar < 10^{9.5}~\msun$, respectively.}
\end{table*}

\begin{figure}
 \centering
 \includegraphics[scale=0.450, viewport=20 0 800 410]{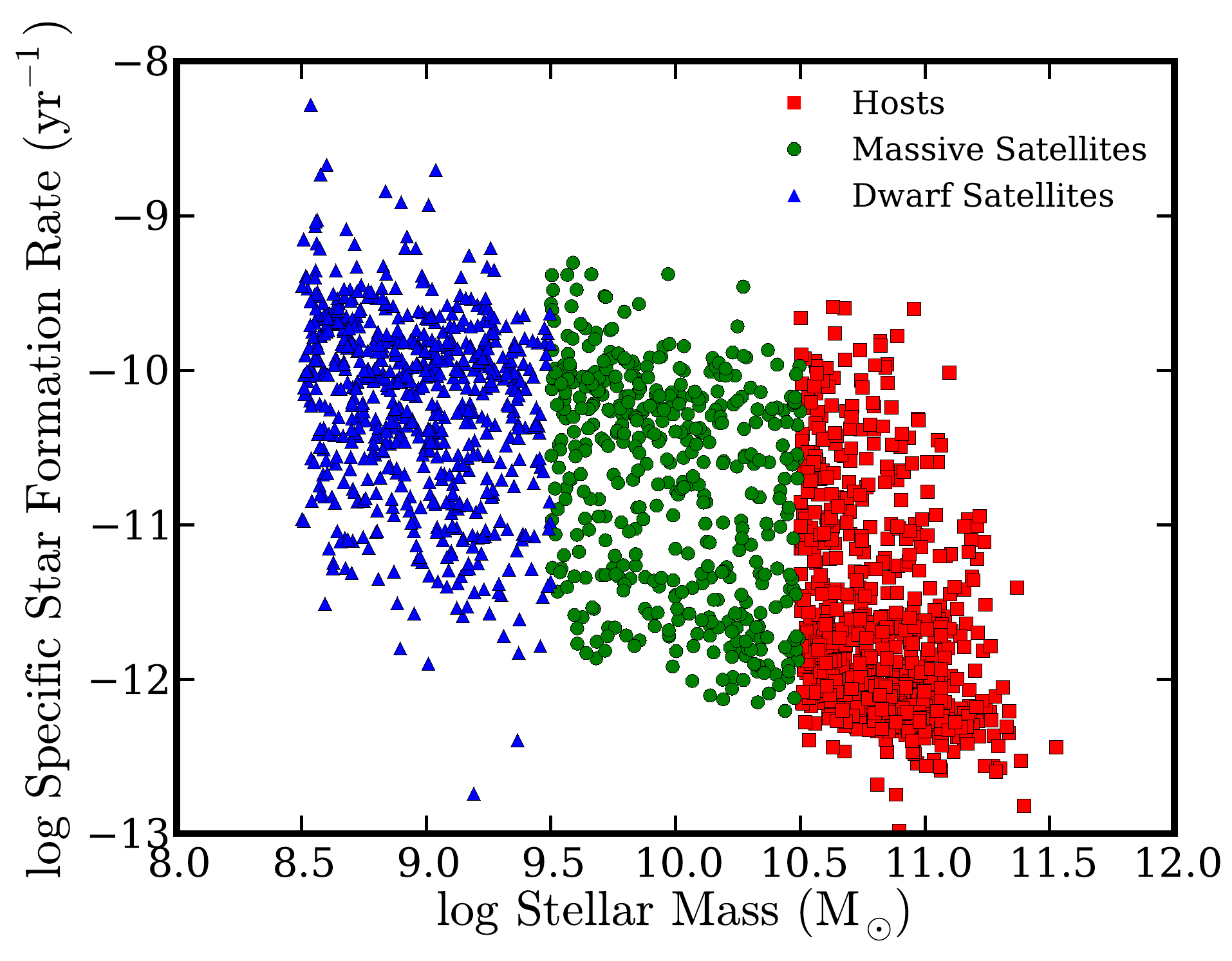}
 \caption{SSFR vs.~stellar mass for our main sample. Plotted are all
   primaries at $z < 0.032$ and their massive ($10^{9.5}~\msun <
   \mstar < 10^{10.5}~\msun$) secondaries. In our main sample, only
   primaries with exactly one massive secondary are considered. Also
   plotted is our dwarf ($10^{8.5}~\msun < \mstar < 10^{9.5}~\msun$)
   secondary sample, which consists only of low-mass secondaries 
   at $z < 0.024$. The dashed line separates objects into passive
   (below line) and star-forming (above line) categories. }
 \label{fig:sample}
\end{figure}

As shown in P14, there are dramatic differences in the effectiveness
of quenching around passive and star-forming Milky Way-size hosts,
such that only massive ($\sim~10^{10}~\msun$) satellites of passive
Milky Way analogs are quenched relative to a stellar mass-matched
sample of isolated field galaxies. Recognizing this dichotomy in
satellite quenching, we divide our sample, including both primaries
and secondaries, according to SSFR, with the division between passive
and star-forming galaxies set to be
\begin{equation} 
\log({\rm SSFR}_{\rm host}) = -0.6\, \log({\rm M}_{\star}) - 5.2 .
\label{eq:thresh}
\end{equation}

This equation is motivated by the established blue cloud/red sequence
bimodality of galaxies in the SDSS \citep{strateva01,baldry04}. This
passivity threshold matches the slope of the stellar mass-SSFR
relation for star-forming galaxies in a self-consistent
way. Figure~\ref{fig:sample} shows where our main and dwarf satellite
samples reside in SSFR-stellar mass space, with the dividing line
between passive and star-forming galaxies overplotted. We note that
our qualitative results are not particularly dependent on our chosen
division between star-forming and passive/quenched, such that other
reasonable choices of quenching definition (e.g.~a quenching threshold
of ${\rm SSFR} < 10^{-10.5}~{\rm yr}^{-1}$) give similar results.

\subsection{Interloper Corrections and Parameter Matching}
\label{sec:interlop}
In this subsection, we discuss two numerical procedures employed to
approximate the distributions of ``true satellite" properties (rather
than simply secondary properties) and to compare samples to
corresponding control samples in a self-consistent way.

In P14, we describe in detail how we connect observations to the
dark-matter only Millennium II Simulation \citep{bk09}. In short, we
wish to use the simulations to disentangle objects that are truly
bound to their hosts from objects that, despite having velocities that
would suggest their being part of the system, are not. This requires
two critical pieces of information: perfect phase-space information
about the objects in question, which the simulations provide, and a
functional model that links dark matter halos to galaxies in the real
Universe. For the latter, we adopt the subhalo abundance matching
(SHAM) prescription of \citet{guo11}. While abundance matching has
difficulty predicting the dark matter halo masses of the local dwarf
galaxy population \citep{bk11, bk12, gk14}, it is very successful in
reproducing the clustering of more massive galaxies, including the
mass range probed by our work \citep{berrier06, conroy06}.

In order to ascertain the distributions and values of parameters for
the true satellite population, we correct for the presence of
interlopers statistically. This is done by taking the cumulative
distribution of a given parameter in the secondary population and
subtracting the distribution of the control sample multiplied by the
probability that a randomly selected galaxy is an interloper (i.e.~$1
- {\rm f}_{\rm purity}$). The result is an unnormalized cumulative
distribution of the parameter in the \emph{satellite} population. We
renormalize so as to produce a well behaved cumulative distribution
function. The equation for the satellite distribution of a parameter,
here F, is given by
\begin{equation}
\label{eq:interloper}
\mathrm{ F_{satellite} = \frac{F_{secondary} - (1-f_{purity})F_{control}}{f_{purity}}} .
\end{equation}
Note that this requires the assumption that all ``impurities" are
isolated interlopers. To ease this assumption, we make the above
calculation using a modified definition of ${\rm f}_{\rm purity}$,
where ${\rm f}_{\rm purity}$ measures the fraction of secondaries in
our sample that are satellites of any host, not necessarily the
identified host.

Often, as in the case of comparisons between satellite and control
samples, we wish to control for distributions of parameters correlated
with those under investigation. For example, to control for mass
dependencies, we match the stellar mass distributions of a given
satellite subsample to a control sample by dividing both subsamples
into bins in the relevant parameter (e.g.~stellar mass). For each bin,
we then randomly select, with replacement, an object from the second
sample for each object in the first sample, yielding two samples of
equal number that are matched on a given parameter. Whenever this is
done, the process is repeated $100$ times.

\section{Satellite Quenching as a Function of System Properties}
\label{sec:results}

In P14, we introduce a parameter to quantify quenching efficiency: the
conversion fraction ($\mathrm{f_{convert}}$), which is defined as the
difference in the quenched fraction between the satellite and control
samples relative to the unquenched fraction of the control sample.  In
other words, let the unquenched fraction $\rm u_{\{sat,control\}}(\log
SSFR=X)$ equal the fraction of satellites or control galaxies with
$\log {\rm SSFR} > {\rm X}$, and the quenched fraction $\rm
q_{\{sat,control\}}(\log SSFR=X)=1-u_{\{sat,control\}}$ be the
fraction of satellites or control galaxies with $\log {\rm SSFR} <
{\rm X}$.  The conversion fraction, ${\rm f}_{\rm convert}$, is then
given by
\begin{eqnarray}
  \label{eq:conv}
\mathrm{f_{convert}} 
&=&\mathrm{\frac{q_{sat}-q_{control}}{u_{control}}}\,.
\end{eqnarray}
In short, ${\rm f}_{\rm convert}$ corresponds to the fraction of
star-forming galaxies that have been quenched upon infall onto the
halos of a given set of hosts. This relies on the assumption that the
properties of the control sample are adequately representative of the
properties of the progenitors of the satellite samples, which is not a
perfect assumption. Most satellites are star-forming, such that they
continue to increase their stellar masses over time, even though they
fell onto a host some time in the past. Other works have used a
similar statistic to our conversion fraction, including \cite{vdb08}
and \cite{peng12}.

Throughout this analysis, errors on the quenched fractions are
computed according to binomial statistics as
\begin{eqnarray}
  \label{eq:error}
\mathrm{\sigma_{q,sample}} 
&=&\mathrm{\sqrt{\frac{(q_{sample})(1 - q_{sample})}{N_{sample}}}}\, ,
\end{eqnarray}
where $\sigma_{\rm q,sample}$ is the error on the quenched fraction of the
sample, ${\rm q}_{\rm sample}$ is the quenched fraction of the sample, and
${\rm N}_{\rm sample}$ is the number of objects in the sample. These errors are
then propagated according to Equation~\ref{eq:conv} to determine the
associated errors on the conversion fractions.


\begin{figure*}
 \centering
 \includegraphics[scale=0.4]{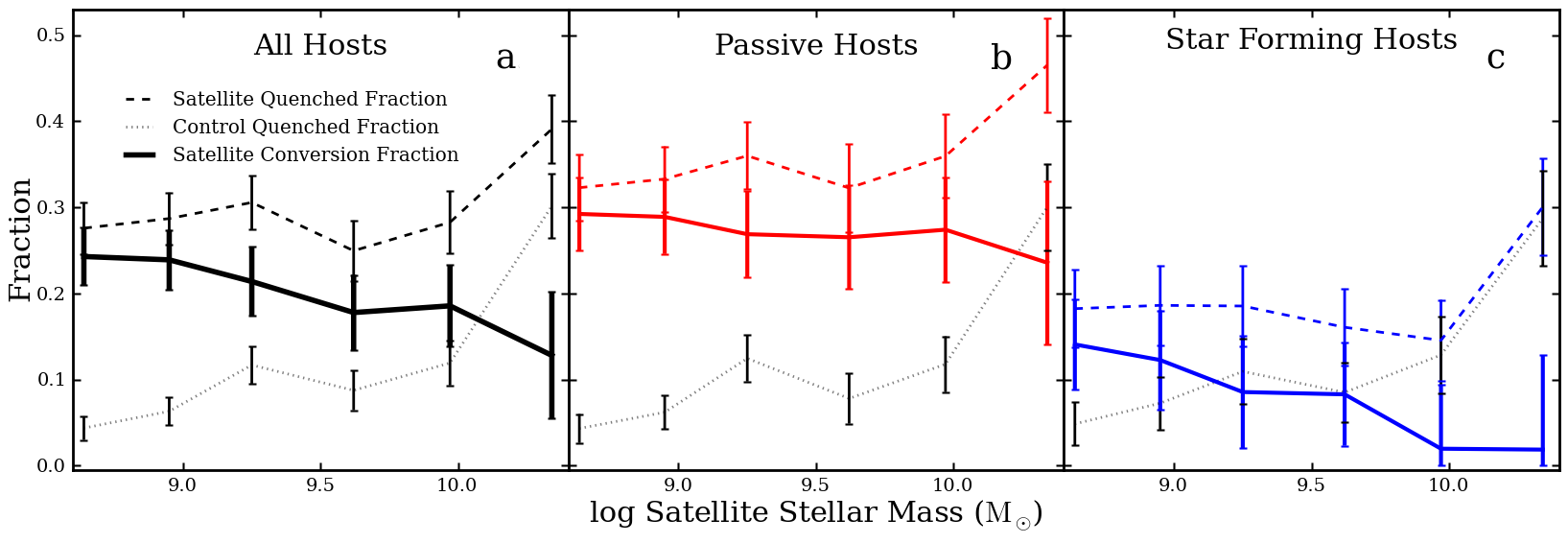}
 \caption{The conversion fraction for satellites of all hosts
   (\emph{left panel}), passive hosts (\emph{center panel}), and
   star-forming hosts (\emph{right panel}) as a function of satellite
   stellar mass in our main sample (\emph{solid lines}). For
   reference, the quenched fraction for the respective satellite
   samples (\emph{dashed lines}) and mass-matched control samples
   (\emph{dotted lines}) are also shown. The conversion fraction is
   independent of stellar mass in the passive host sample, while
   exhibiting a negative correlation with satellite mass in the
   star-forming host sample, such that less-massive satellites of
   star-forming hosts are more likely to be quenched upon infall. }
 \label{fig:sat_mass}
\end{figure*}


\subsection{Dependence on Satellite Mass}

Considering both our main and dwarf subsamples, our full sample of
satellite galaxies spans two orders of magnitude in stellar mass,
$10^{8.5} < \mstar/\msun < 10^{10.5}$. To explore the dependence of
quenching on stellar mass, we separate our satellite sample into six
distinct bins, by independently dividing the main and dwarf subsamples
according to the $33^{\rm rd}$ and $67^{\rm th}$ percentiles in
stellar mass. The resulting six independent bins are bounded in
stellar mass space by $10^{8.5}$, $10^{8.80}$, $10^{9.11}$,
$10^{9.5}$, $10^{9.79}$, $10^{10.14}$, and $10^{10.5}~\msun$. In
Figure~\ref{fig:sat_mass}a, we show the passive fraction for the
satellite samples in each stellar mass bin (dashed line) along side
that of the control samples of isolated galaxies (dotted line) with
the same stellar mass distribution (i.e.~``mass-matched''). As
highlighted by previous studies of satellite galaxies in the local
Universe, predominantly in more-massive halos, we find that satellites
are preferentially passive relative to the field population
\citep[e.g.][]{weinmann06, Tollerud11, geha12, wang12}.

Following Equation~\ref{eq:conv}, the measured passive fractions for
the satellite and control samples yield a conversion fraction as a
function of satellite mass that weakly increases with decreasing
stellar mass, such that less-massive satellites are slightly more
likely to be quenched upon infall to a Milky Way-like halo (see solid
line in Fig.~\ref{fig:sat_mass}a). Within the errors, however, the
measured conversion fractions are largely consistent with no
dependence on satellite stellar mass at $10^{8.5}~\msun< \mstar <
10^{10.5}~\msun$. Using repeated Monte Carlo resampling and assuming
the stated errors are normally distributed,\footnote{Our samples are
  large enough that the assumption of normally-distributed errors is
  valid, typically on the order of n = 60 - 100.} we estimate the slope of the relation between conversion
fraction and $\log\mstar$ to be $-0.069 \pm 0.036$, marginally
inconsistant with no correlation and consistent with a slight
anti-correlation.

As shown by P14, the conversion fraction (and thus efficiency of
satellite quenching) varies significantly between passive and
star-forming Milky Way-like hosts, such that massive ($\sim \!
10^{10}~\msun$) satellites are only quenched around passive
hosts. Given this observed dichotomy of massive satellite quenching,
we compute the passive and conversion fractions as a function of
satellite stellar mass for passive and star-forming hosts separately
(see Figure~\ref{fig:sat_mass}b,c). At all satellite stellar masses
probed, passive hosts are more effective at quenching than
star-forming hosts. Moreover, we again find that at high satellite
masses, passive hosts are the sole drivers of satellite quenching,
with a conversion fraction for satellites of passive hosts of roughly
$30\%$ relative to nearly $0\%$ around star-forming systems.
Remarkably, across the entire range of satellite stellar masses
studied, this moderate quenching efficiency (of $\sim \! 30\%$) is
relatively independent of satellite mass for passive hosts. Using the
Monte Carlo method described above, we find a slope in the conversion
fraction-satellite stellar mass relation of $-0.027 \pm 0.048$,
consistent with no dependence of quenching efficiency on satellite
stellar mass for passive Milky Way-like hosts.  Conversely, we find
weak evidence for a modest increase in the conversion fraction at
lower stellar masses around star-forming systems, with a slope of
$-0.089 \pm 0.058$. This suggests that the weak negative slope seen
with passive and star forming hosts taken together is driven by the
anti-correlation in quenching efficiency and satellite stellar mass
found around star-forming hosts.

\begin{figure*}
 \centering
 \includegraphics[scale=0.40]{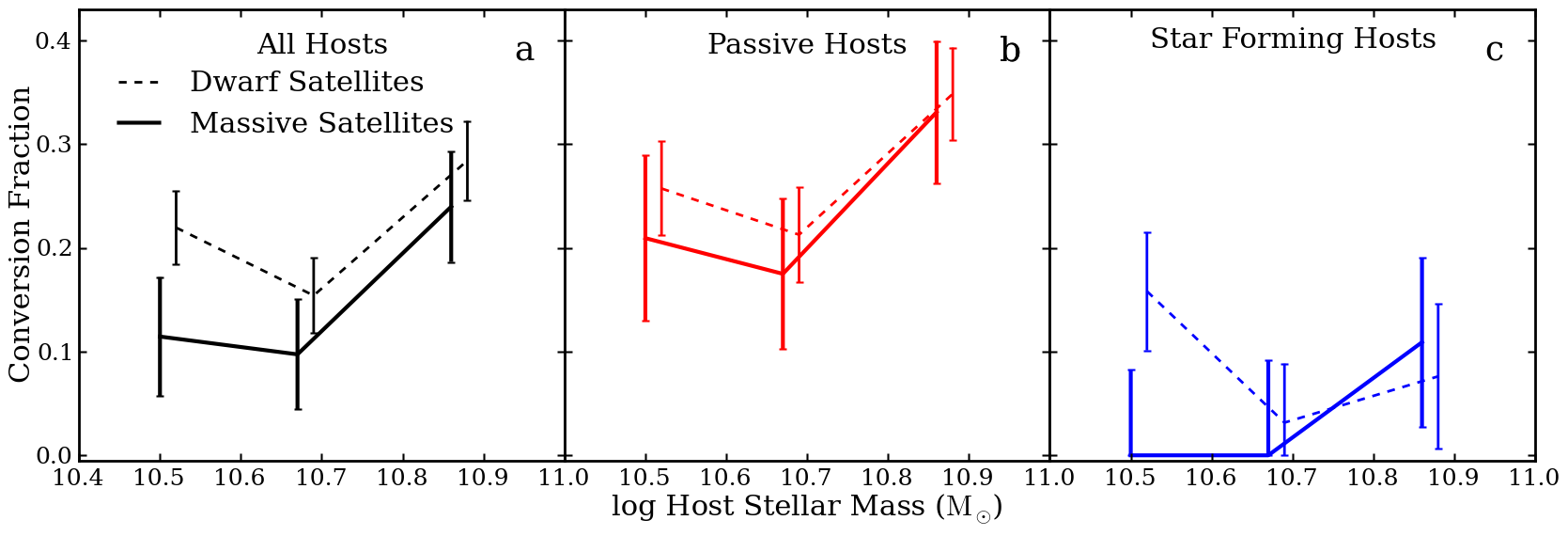}
 \caption{Conversion fractions for satellites in our main sample
   (\emph{solid lines}) and dwarf sample (\emph{dashed lines}) as a
   function of the stellar mass of their host, divided according to
   the star-forming properties of the host. Left panel: all hosts,
   center panel: passive hosts, right panel: star-forming hosts. We
   find little evidence for a dependence of satellite quenching
   effiency on host stellar mass.}
 \label{fig:host_mass}
\end{figure*}

\subsection{Dependence on Host Mass}

It is natural to expect that the efficiency of satellite quenching
correlates with the stellar mass of the host. For example, galaxies of
higher stellar mass tend to live in more massive dark matter halos
\citep{mandelbaum06, conroy07, behroozi12, mcgaugh12, moster13,
  Miller11, miller14}, which may quench satellites more
effectively. This trend is observed on large scales, with clusters
hosting significantly higher fractions of passive galaxies relative to
comparable field samples \citep[e.g.][]{dressler80, postman05}. With
this thought in mind, we study the dependence of quenching efficiency
on host stellar mass within our sample of less-extreme, lower-mass
halos.  Hosts are divided into three mass bins bounded by the $33^{\rm
  rd}$ and $67^{\rm th}$ percentiles of the host stellar mass
distribution, which correspond to stellar masses of $10^{10.69}~\msun$
and $10^{10.86}~\msun$, respectively.

To examine how quenching efficiency depends on the stellar mass of the
host, we examine the satellites in the main and dwarf sample
separately, as shown in Figure \ref{fig:host_mass}, and divide the
hosts according to their status as passive or star-forming. This gives
us four independent subcategories that we can use to examine the host
stellar mass dependence of quenching efficiency: massive satellites of
passive hosts, massive satellites of star-forming hosts, dwarf
satellites of passive hosts, and dwarf satellites of star-forming
hosts. Of the four subsamples, all but the dwarf satellites of
star-forming hosts exhibit a similar correlation between ${\rm f}_{\rm
  convert}$ and host stellar mass: constant quenching efficiency in
the two lower host mass bins, with a slight increase in quenching
efficiency in the highest mass bin. We again use Monte Carlo resampling to test whether the slope of the conversion fraction-host stellar mass relation is consistent with zero in each subsample --- that is, wether or not the increase in quenching
efficiency at high host stellar mass is statistically significant. In
all cases, we find that the observed quenched efficiencies are largely
consistent with no dependence on host mass, such that the measured
slopes of the conversion fraction versus $\log(\mstar)$ relation are
consistent with zero at $<2\sigma$.

The subsample that differs significantly from the other three
categories, dwarf satellites of star-forming hosts, sees an increase
in quenching efficiency around the lowest mass hosts.  We showed in
the previous subsection that lower mass satellites of star-forming
hosts are more likely to be quenched; however, the increase in the
efficiency with which dwarf satellites are quenched in low-mass
star-forming hosts is likely not driven by an decrease in the
characteristic stellar mass of satellites of such hosts. A
Kolmogorov-Smirnov (KS) test fails to show that the stellar masses of
the objects in the three dwarf satellite/star-forming host bins are
drawn from different underlying distributions ($p_{lower,middle} =
0.71$, $p_{lower,upper} = 0.37$). While the behavior of dwarf
satellites of star-forming galaxies does seem odd, the data are
consistent with no dependance in conversion fraction on host stellar
mass.

\begin{figure*}
 \centering
 \includegraphics[scale=0.34]{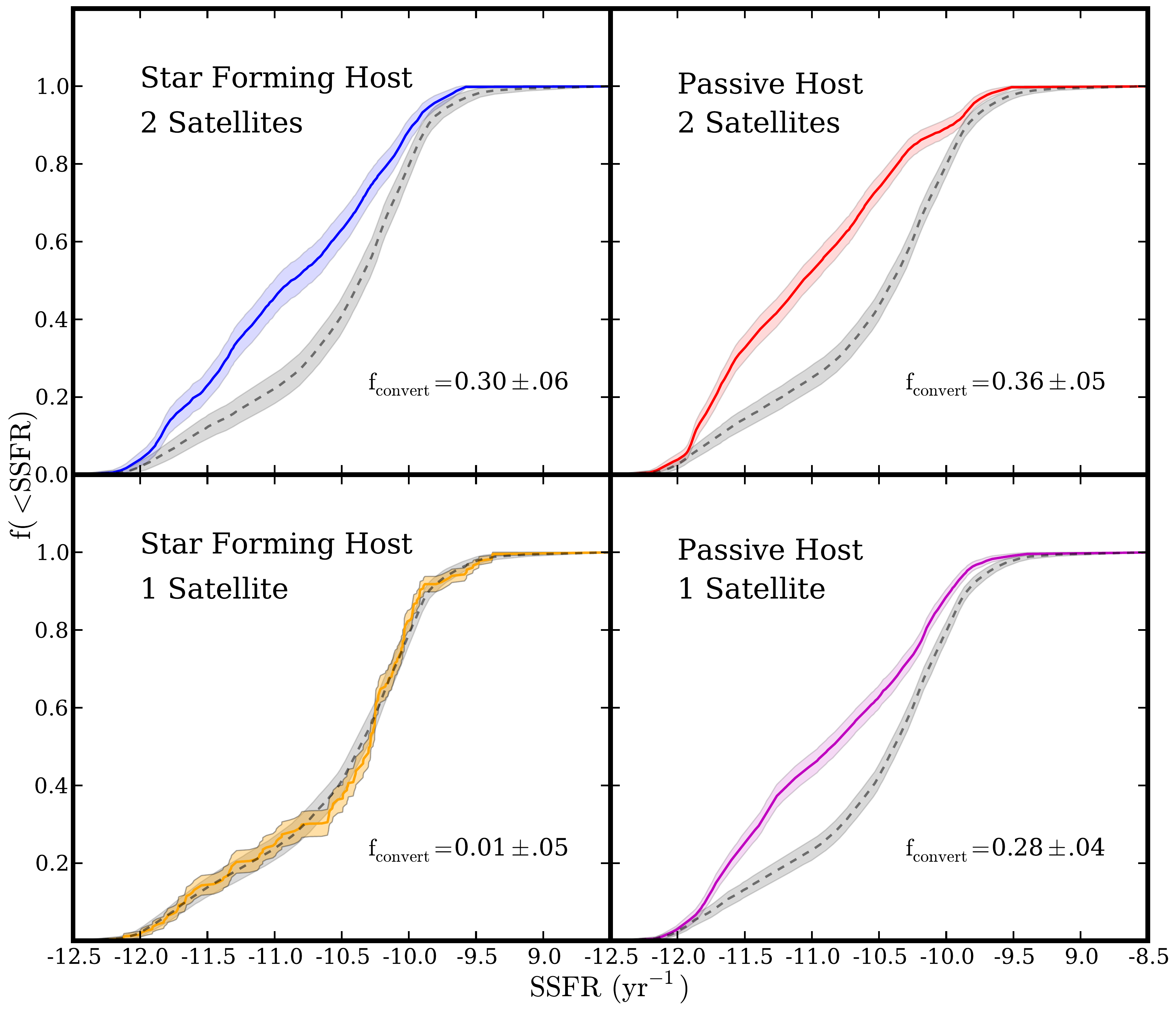}
 \caption{Cumulative distributions of specific star formation rate for
   massive satellites of passive (right column) and star-forming (left
   column) hosts with exactly one (bottom row) and exactly two (top
   row) satellites. The black dashed lines show the corresponding
   distributions for stellar mass-matched samples of isolated field
   galaxies. Colored and grey shaded regions correspond to $1\sigma$
   binomial errors for the satellite and control samples,
   respectively.  Hosts with two massive satellites are more effective
   at quenching those satellites than corresponding hosts with only
   one satellite, and star-forming hosts with two satellites have a
   non-zero quenching efficiency, breaking the dichotomy of quenching
   presented in P14.}
\label{fig:2sats}
\end{figure*}

\subsection{Dependence on Satellite Number}
\label{sec:satnum}

As discussed in \S\ref{sec:sim}, we have to this point restricted our
analysis to hosts with exactly one massive satellite, as this
preferentially selects Milky Way-like systems in lieu of more massive
dark matter halos (see Fig.~\ref{fig:mtop}). Here, we relax this restriction on the number of massive satellites and consider separately hosts that have exactly one massive satellite and hosts with more than one massive satellite. These categories
represent hosts with different characteristic virial masses. As shown
in Fig.~\ref{fig:mtop}, we find that hosts with two massive satellites
live in systematically more massive dark matter halos based on
comparison to the MS-II simulation.\footnote{The median halo mass for
  isolated hosts with two massive secondaries is approximately $60\%$
  larger than that for isolated systems with only one massive
  secondary in the MS-II simulation.}
For ease of comparison with P14, we adopt a simplified threshold for
quenching, by which satellites will be considered quenched if their
SSFR is below $10^{-11}~{\rm yr}^{-1}$. In Figure~\ref{fig:2sats}, we
plot the cumulative distribution of specific star formation rates for
massive satellites around passive and star-forming hosts separately,
matching the stellar mass distributions of all four host samples. For
hosts with only one massive satellite, the dichotomy of quenching
discovered in P14 is readily apparent, with satellites of passive
hosts more than twice as likely to be quenched than a stellar
mass-matched field sample and satellites of star-forming hosts
indistinguishable from their field counterparts. For systems with
multiple massive satellites, however, the behavior is
different. Passive hosts with two satellites still quench their
satellites more efficiently than star-forming hosts with two
satellites, but star-forming hosts now exhibit non-zero quenching
efficiency, possibly pointing to a trend in quenching efficiency with
host halo mass.

In light of the evidence that more massive halos quench their
satellites more effectively, we might hypothesize that there exists a
critical halo mass threshold above which a host will quench its
satellites extremely efficiently. This could be considered a potential
limiting case of quenching scenarios, whereby in a two-satellite
system the conditional probability of finding a satellite quenched
given that its partner is quenched is near unity. An alternate
limiting case might be that the probabilities of finding either
satellite quenched are independent of each other. How well the data
conform to either limiting case can inform quenching models, and by
examining our sample host-by-host, we can examine how the satellites
are distributed among the hosts.

\begin{figure*}
 \centering
 \includegraphics[scale=0.44]{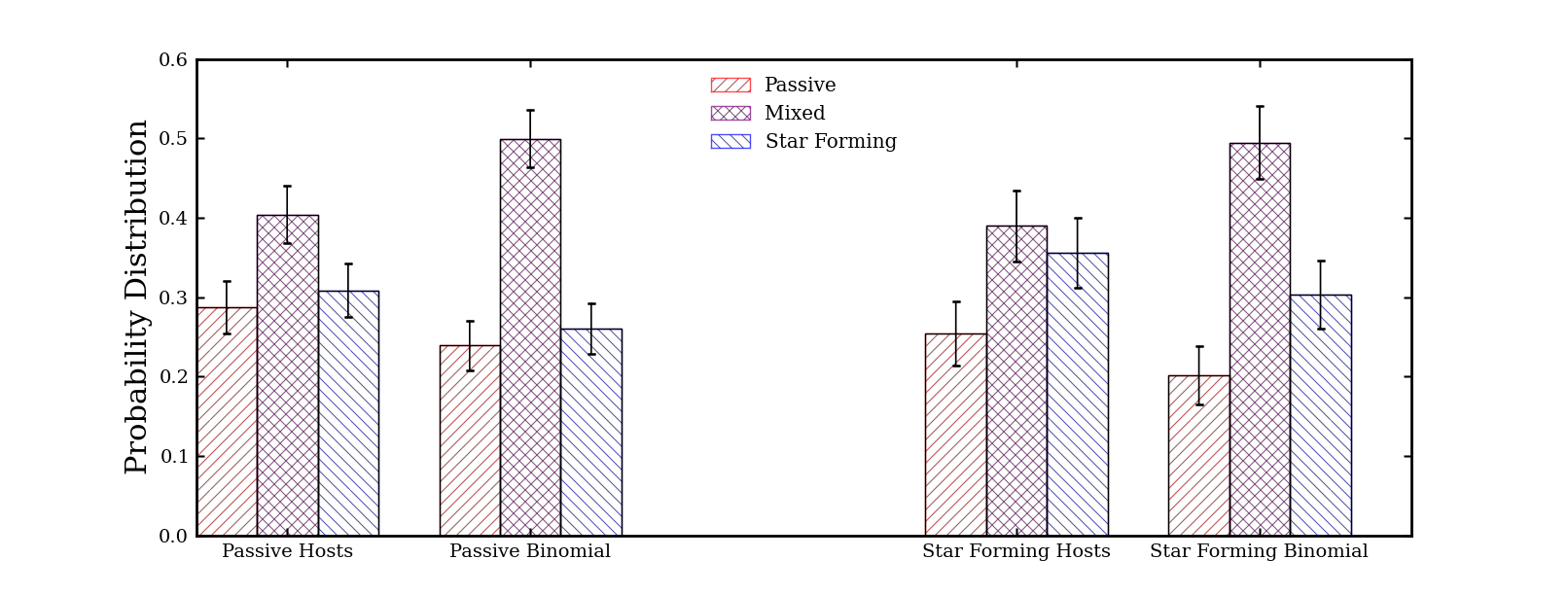}
 \caption{The relative frequency of matched or mis-matched satellite
   properties in two satellite systems for passive (\emph{left}) and
   star-forming (\emph{right}) hosts. Shown separately are the
   frequency of matched passive pairs (red bars), matched star-forming
   pairs (blue bars) and mixed pairs (i.e.~one passive satellite and
   one star-forming satellite, purple bars). Also plotted are binomial
   distributions calculated from the red fraction of satellites in the
   parent passive and star-forming host categories, representative of
   the limiting case where satellites are randomly assigned to
   hosts. In systems with two satellites, having one passive satellite
   corresponds to a slightlyhigher likelihood of the second satellite
   also being passive and vice versa for star-forming satellites. }
\label{fig:2satbar}
\end{figure*}

Considering each two-satellite system individually, we ask whether the
satellites are matched or unmatched in their star-forming properties.
Figure~\ref{fig:2satbar} shows the frequency of passive, star-forming,
and mis-matched satellite pairs around passive and star-forming
hosts. For comparison, we also plot the binomial distributions for
both samples --- i.e.~the expectation from randomly drawing satellites
from the population of all satellites. In the observational sample,
mis-matched pairs are better represented than the matched pairs of
passive or star-forming satellites. While the incidence of mis-matched
pairs is lower than the binomial distribution and the frequencies of
matched passive and star-forming pairs are higher than the binomial
distribution, the disagreement is fairly minor. For a system with two
massive satellites, having one satellite be passive corresponds to a
slightly higher probability of its partner being passive and vice
versa for star-forming satellites. We note that the distributions are
very similar between the two host types, and despite statistically
signifiant deviation from the binomial distribution, the distribution
of observed pairs is much better modeled by a binomial distribution
than the limiting case of all satellites being found in matched pairs.

\begin{figure*}
 \centering
 \includegraphics[scale=0.24]{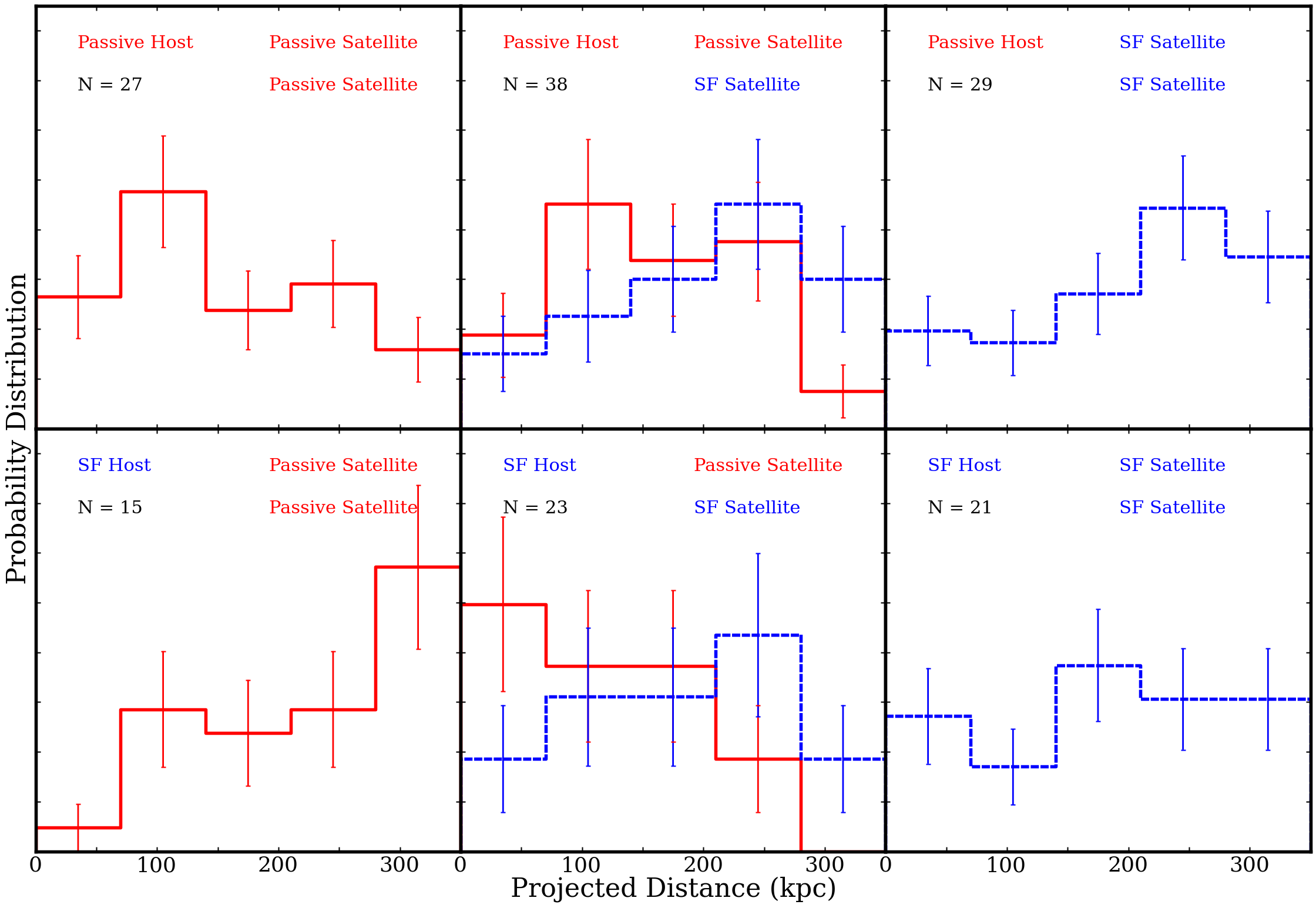}
 \caption{Radial distributions of massive satellites for passive
   (\emph{top row}) and star-forming (\emph{bottom row}) hosts with
   exactly two massive satellites. In each row, we divide the
   subsamples into passive satellite pairs (\emph{left column}), mixed
   satellite pairs (\emph{center column}), and star-forming satellite
   pairs (\emph{right column}). In each case, passive satellite
   distributions are represented by a solid red line and star-forming
   satellite distributions by a dashed blue line. There is an overall
   trend for passive satellites to be more centrally concentrated than
   their star-forming counterparts. The overabundance of passive
   satellites at large projected distances in the bottom left panel
   may be a result of mis-identifying the host galaxy.}
\label{fig:2satrad}
\end{figure*}

Satellite galactocentric distance can provide information about the
environment in which it lives, such as what type of CGM it is embedded
in or how long it has been interacting with its host. With that in
mind, we examine the projected distance distribution of satellite
galaxies in systems with exactly two massive satellites. In
Figure~\ref{fig:2satrad}, we examine the radial distribution of
satellites in systems with two massive satellite, grouping
the satellites according to their star-forming properties as well as
that of the host. The top and bottom sets of three panels show the
distributions for satellites of passive and star-forming hosts,
respectively.

Of particular interest is the difference between the radial
distributions of passive satellites compared to their star-forming
counterparts. We find that passive satellites are more likely to be
found at small projected radii, while there is a corresponding
overabundance of star-forming satellites in the outer regions of the
halo. This trend is apparent in the mixed satellite cases, and is
readily seen upon comparing passive and star-forming satellites across
the various host subsamples. A Kolmogorov-Smirnov (KS) test rejects
the null hypothesis that the distributions of projected distances are
identical between star-forming and passive satellites ($p =
0.004$). This trend is particularly apparent in the cases where one
satellite is star-forming and one satellite is passive (middle panels
of Fig.~\ref{fig:2satrad}); for both passive and star-forming hosts,
the star-forming satellites are more likely to be found in the outer
regions of the host halo relative to the passive satellites.

The one category that seems not to follow the trend is the category of
star-forming hosts with paired passive satellites. Here we find,
despite low statistics, that satellites are mostly found at large
projected distances. One possible interpretation of this result is
that a relatively high portion of these objects were not
environmentally quenched by their host, but rather quenched in the
field. We might expect that if the satellites had been environmentally
quenched at a similar rate to the other categories, they would be
found at projected distances more in line with those seen in the other
categories. Alternately, the apparent inversion of the radial trend
seen in the other subsamples could be the result of host
misidentification. In situations where the hosts and satellites have a
nearly one-to-one mass ratio, if the true host of the system is
misidentified as a distant satellite, the unexpected case of a host
having a distant passive satellite would indeed be the expected case
of a host having a distant star-forming satellite. Of the 30 objects
identified as satellites in this subsample, 12 have projected
distances greater than $250~\kpc$. Of these, 5 are within a factor of
three of their host in stellar mass, comprising 4 of the 15
systems. If these four systems are disregarded, the radial
distribution of passive satellites becomes flat, more in line with the
other subsamples of passive galaxies.

\section{Discussion}
\label{sec:discuss}

In this paper, we have examined the dependence of satellite quenching
on the mass of the satellite galaxy, the mass of the host galaxy, and
the multiplicity of satellites in the system. For our host and
satellite sample, we show that satellite quenching is not strongly
correlated with satellite mass and potentially increases with
increasing host virial mass, as best shown by an increase in
conversion fraction in systems with two massive satellites. We also
demonstrate that within such systems, passive and star-forming
satellites are found around hosts roughly with frequencies described
by a binomial distribution, implying little correlation between the
properties of satellite pairs. In the following section, we discuss
these results in comparison to previous studies in the local Universe
and highlight the implications of these findings on models of galaxy
evolution.

\subsection{The Mass Dependence of Satellite Quenching}

Across a broad range in stellar mass, $10^{8.5}~\msun < \mstar <
10^{10.5}~\msun$, we find little evidence for a correlation between
conversion fraction (i.e.~the efficiency of satellite quenching) and
satellite stellar mass, with Milky Way-like hosts quenching
$\sim~\!~20\%$ of infalling satellites on average. This relatively
inefficient quenching is strongly correlated with host properties,
such that the satellites of passive hosts, at all masses probed, are
more likely to be quenched than their counterparts around star-forming
hosts. Moreover, for the most massive satellites
($\sim~\!~10^{10}~\msun$), quenching is entirely driven by the halos
of passive hosts, with massive satellites of star-forming hosts
indistinguishable from the field population --- a confirmation of the
dichotomy of satellite quenching shown in P14.  At lower satellite
stellar masses ($\sim~\!~10^{8.5}~\msun$), however, we measure a
conversion fraction of $\sim~\!~15\%$ for satellites of star-forming
hosts. This breaks the strong dichotomy of satellite quenching
observed at higher masses, though the mass dependance of satellite
quenching efficiency for satellites of star-forming hosts is mild.

The lack of significant correlation between satellite stellar mass and
conversion fraction is surprising. In accordance with subhalo
abundance matching, lower stellar mass galaxies tend to occupy lower
virial mass (and thus less dense) halos. Given their shallower
potential wells, these galaxies would therefore be expected to lose
their gas more easily and become quenched, thereby yielding a higher
conversion fraction at lower satellite masses. However, this is not
observed.

One possibility is that the expected trend for low-mass galaxies to
lose their gas (and thus quench) more easily may be balanced out by a
tendency for these systems to possess larger gas reservoirs. Under a
scenario where hot gas is stripped from a satellite upon infall but
cold gas is retained (i.e.~``strangulation"), galaxies with higher
cold gas fractions would thereby take longer to use up their gas and
ultimately quench. According to observations of atomic hydrogen in
local star-forming galaxies, lower stellar mass systems are generally
found to have higher atomic gas fractions and longer atomic depletion
timescales (${\rm SFR}/{\rm M}_{\rm HI}$) than their more massive
counterparts \citep{skillman03, geha06, leroy08, schiminovich10}.
Furthermore, recent studies of molecular gas, which more closely
traces ongoing star formation \citep{wong02, Kennicutt07, bigiel08},
also show increasing cold gas fractions at lower stellar masses
\citep{saintonge11, boselli14}. Altogether, the overall trend for the
progenitors of low-mass satellite to have high cold gas fractions and
correspondingly long depletion times may serve to counteract the
tendency for low-mass satellites to lose gas easily, resulting in an
overall non-dependence of conversion fraction on satellite stellar
mass. While the efficiency of satellite quenching is observed to be
relatively independent of satellite stellar mass, we do find a
significant increase in the conversion fraction, ${\rm f}_{\rm
  convert}$, with increasing host halo mass. In particular, we find a
greater satellite quenching efficiency for host systems with two
massive satellites relative to those with only one massive satellite
(see Fig.~\ref{fig:2sats}), where comparison to the MS-II simulation
shows that systems with more massive satellites are preferentially
biased towards greater dark matter virial masses (see
Fig.~\ref{fig:mtop}). Our measurements of stacked satellite velocity
dispersions confirm that hosts with two massive satellites
preferentially reside in more massive halos. As in P14, we stack the
line-of-sight velocity distributions for satellites of passive and
star-forming hosts at fixed host stellar mass, so as to measure the
velocity dispersion as a proxy for the characteristic virial mass of
the respective host populations. Passive hosts with one satellite have
a satellite velocity dispersion of $165 \pm 11~\kms$, while passive
hosts with two satellites have a satellite velocity dispersion of $197
\pm 13~\kms$, such that passive hosts with two satellites live in more
massive dark matter halos than passive hosts with one satellite. Likewise,
star-forming hosts with a single massive satellite yield a stacked
satellite velocity dispersion of $148 \pm 14~\kms$, whereas
star-forming hosts with two massive satellites have a satellite
velocity dispersion of $189 \pm 16~\kms$, implying the same about
star-forming hosts.

The apparent correlation between satellite quenching efficiency and
host halo mass may also explain the increased prevalence of quenched
satellites around passive hosts relative to their star-forming
counterparts. As shown in Fig.~\ref{fig:2sats}, the conversion
fraction for satellites of passive Milky Way-like hosts, in both the
one- and two-satellite cases, exceeds that measured for satellites of
star-forming hosts (see also Fig.~\ref{fig:sat_mass}). The comparison
of stacked satellite velocity dispersions for these samples shows that
even at fixed satellite number, as well as fixed stellar mass, passive
hosts preferentially live in more massive dark matter halos. Thus, as
discussed by P14, a quenching efficiency that depends on host halo
mass in concert with the preference for passive hosts --- at a given
stellar mass --- to reside in more massive halos directly explains the
increased prevalence of quenched satellites around passive hosts.

Beyond dynamical tracers of halo mass, host stellar mass is also
expected to track halo mass on average, with more massive host
galaxies typically residing in more massive halos
\citep[e.g.][]{moster10, miller14}. Yet, while we do find mild
evidence for an increase in quenching efficiency at high host stellar
mass, our results are largely consistent with no correlation between
conversion fraction and the stellar mass of the host. One possible
explanation for this lack of observed correlation between satellite
quenching and host stellar mass is that our sample spans a relatively
narrow range in host stellar mass, such that we cannot resolve a clear
stellar mass-halo mass relation --- i.e.~the stellar mass-halo mass
relation may be largely dominated by scatter over the range in halo
masses probed by our hosts. To test this possibility, we again stack
the satellites of the hosts in each host stellar mass bin and measure
the velocity dispersions of the massive satellite populations. We find
no trend in satellite velocity dispersion with increasing host stellar
mass: the lowest bin in host stellar mass has a velocity dispersion of
$163 \pm 10~\kms$, the middle bin has a velocity dispersion of $162
\pm 11~\kms$ and the upper bin has a velocity dispersion of $155 \pm
9~\kms$. This suggests that the evidence we find for a strong
dependence of quenching efficiency on host halo mass and for weak to
no dependence of quenching efficiency on host stellar mass do not
necessarily contradict each other and are consistent with a picture
where higher virial mass hosts are more likely to quench their
satellites.

As discussed in P14, the dependence of quenching efficiency on host
halo mass may reflect the preference for more massive dark matter
halos to harbor hot gas coronas, which are then able to quench
infalling satellite galaxies via ram-pressure stripping. Studies of
gas accretion onto dark matter halos indicate that there is a
transition in the dominant accretion mode at a halo mass of roughly
${\rm a~few} \times 10^{12}~\msun$. Infalling gas is shock-heated at
the virial radius in halos above this threshold, while cold gas
reaches a smaller radius, possibly falling all the way to the galaxy,
in less-massive halos \citep[e.g.][]{binney77, rees77, birnboim03,
  keres05, keres09, stewart11}. Models of galaxy formation in which
quenching only occurs above this critical halo mass show significant
promise in reproducing the observed dependence of the galaxy quenched
fraction on stellar mass and environment at $>~\!\!\!~10^{9.5}~\msun$
\citep{gabor14}. Our results, as presented in Fig.~\ref{fig:2sats}, are largely
consistent with this picture of gas accretion and stripping, such that
star-forming hosts with one massive satellite preferentially reside in
halos below the critical halo mass and passive hosts with one massive
satellite as well as all hosts with two massive satellites inhabit
more massive halos with established hot coronas. In the less massive
halos, the satellite galaxies largely mirror the field population,
while roughly $30\%$ of infalling satellites are quenched in more
massive systems, potentially due to the presence of a hot halo.

It should be noted, however, that models such as these, which depend
on interactions with host circumgalactic media to drive satellite
quenching, may struggle to reproduce the observed correlation between
the star-forming properties of massive galaxies and that of lower-mass
systems located at distances of several virial radii
\citep[e.g.][]{kauffmann13, Wetzel14}. Alternatively, work by
\cite{Hearin14} presents a picture where the importance of intra-halo quenching mechanisms is overstated, and large scale conformity, i.e. two-halo effects, is important.

While clear evidence is found for a correlation between quenching
efficiency and halo mass, the observations of two-satellite systems
are not -- at first glance -- entirely consistent with this picture.
As shown in Fig.~\ref{fig:2satbar}, the star-forming properties of
satellites are consistent with being randomly drawn from the parent
population, such that if one massive satellite is quenched in a system
then the second satellite is only marginally more likely to also be
quenched. In a scenario where quenching is driven entirely by host
halo mass, one might expect that quenched satellites would be
preferentially found within a particular subset of halos (e.g.~within
those more massive halos with a CGM capable of stripping an infalling
satellite). This reasoning, however, assumes that satellites quench on
a reasonably short timescale, such that there is a relatively small
chance of observing a star-forming satellite within a halo capable of
quenching it. Recent studies of quenching timescales for low-mass
satellites suggest that this assumption is flawed, with satellites
with a stellar mass of $\sim \! 10^{10}~\msun$ estimated to quench
$\sim \! 6~{\rm Gyr}$ after infall \citep{delucia12, wetzel13a,
  wheeler14}. Adopting this quenching timescale within the MS-II
simulation, where infall times are known, and assuming that all
systems with two massive satellites are above the critical halo mass,
we are able to precisely reproduce the distributions of matched and
mis-matched satellite pairs shown in Figure~\ref{fig:2satbar}, under
the assumption that only subhalos accreted more than $5$ Gyr ago are
quenched. Additional models could potentially be constrained with
quenching timescales that vary with host mass and a critical halo mass
above which a host can quench and below which it can not.

\begin{figure*}
 \centering
 \includegraphics[scale=0.35]{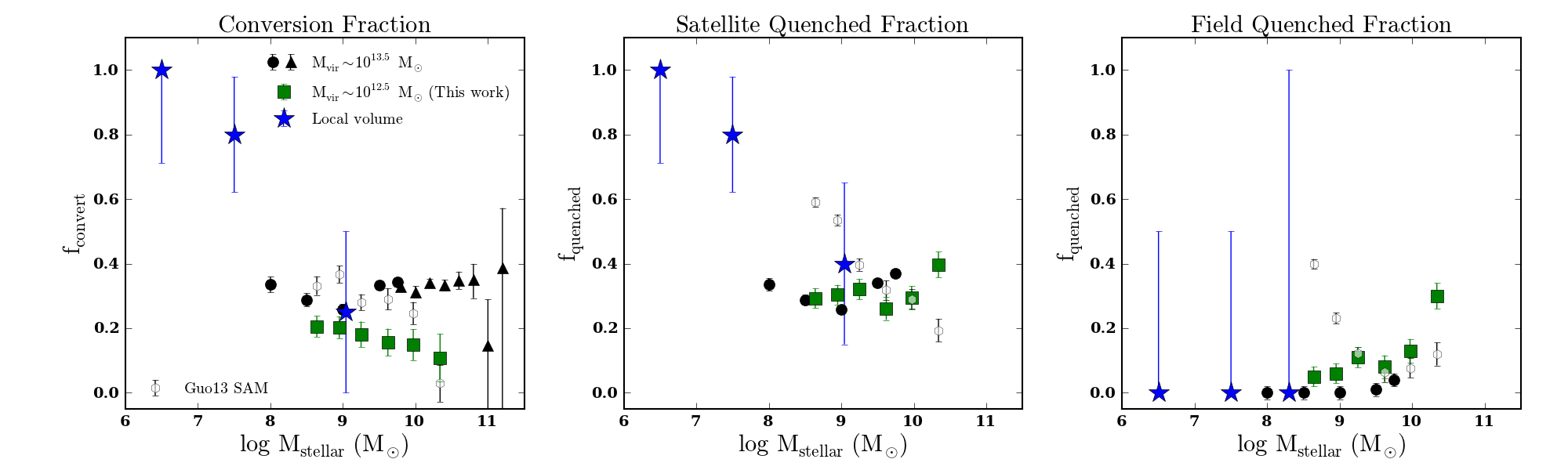}
 \caption{Conversion fractions (left panel), satellite quenched
   fractions (center panel), and field quenched fractions (right
   panel) for our sample of Milky Way-like systems (${\rm M}_{\rm vir}
   \sim 10^{12.5}~\msun$; green squares). For comparison, we show
   various samples of local galaxies including group-scale systems
   (${\rm M}_{\rm vir} \sim 10^{13.5}~\msun$, black circles:
   \citealt{geha12,wheeler14}; black triangles: \citealt{wetzel13a}),
   the Local Group/Local Volume (blue stars) and simulated galaxies
   from the \citet{guo13} SAM (open grey hexagons). At stellar masses
   greater than $10^{8}~\msun$, conversion fractions show little to no
   dependance on satellite stellar mass at fixed host halo mass. At
   fixed satellite stellar mass, hosts with greater virial masses
   quench their satellites more effectively. A significantly greater
   fraction of Local Group satellites are quenched than higher mass
   satellites, indicating a potential critical mass scale for
   satellite quenching at $\sim 10^{8}~\msun$. The differences between
   the SAM and the observational data is likely driven by
   overpredicting the field quenched fraction at these masses. Note
   that we exclude likely ``backsplash galaxies" (i.e.~galaxies
   believed to have previously interacted with their host, namely
   Cetus) from the Local Field sample.}
 \label{fig:LG_sats}
\end{figure*}

\subsection{Comparison to Previous Studies}

Using data drawn from the SDSS, several recent studies similarly
conclude that satellite quenching efficiency is largely independent of
satellite stellar mass at $10^{8} \lesssim \mstar/\msun \lesssim
10^{11}$ \citep{geha12, wetzel13a, wheeler14}. In particular,
\citet{wetzel13a} employ the group catalog of \citet{yang05, yang07}
to study the frequency of quenched satellites at $> 10^{9.7}~\msun$,
showing that $\sim 30\%$ of massive satellites are quenched upon
infall to a host halo. Using the data of \citet{geha12} to push to
lower satellite masses, \citet{wheeler14} extend this work by studying
the quenching efficiency around local hosts with stellar masses of
$>10^{10.4}~\msun$, finding a comparable quenching efficiency. Figure~\ref{fig:LG_sats} shows our measurements of (\emph{i})
satellite conversion fraction, (\emph{ii}) satellite quenched
fraction, and (\emph{iii}) field quenched fraction as function of
satellite stellar mass for Milky Way-like hosts (i.e.~our ``main
sample''). For comparison, we include complementary results from
\citet{wheeler14} and \citet{wetzel13a} along side corresponding
measurements for the Local Group and the \citet{guo13} semi-analytic
model (SAM). Both the \citet{wheeler14} and \citet{wetzel13a} studies find higher
conversion fractions at all stellar masses. This relative increase in
quenching efficiency is likely driven by variations in the sample
selection that lead to significant differences in the typical halo
masses probed. As shown by \citet{wheeler14}, the typical host in the
\citet{geha12} sample has a halo mass of $\sim 10^{13.5}~\msun$,
comparable to that of the group catalog employed by \citet{wetzel13a}.
Our methodology differs in that we apply isolation criteria so as to
restrict our analysis to $10^{12}~\msun$ halos. This places our hosts
at systematically lower virial masses than the above works, which in
turn biases our results towards environments of lower quenching
efficiency. 

Our work, taken together with the complementary results described
above, paints a picture of relatively low satellite quenching
efficiency ($\sim 30\%$) that is independent of satellite stellar mass
across an impressively broad range in mass ($10^{8} < \mstar/\msun <
10^{11}$). At stellar masses of $>10^{8}~\msun$, observations of the
Local Group are remarkably consistent with these results. Among the
massive satellites of the Milky Way and M31, the LMC, SMC, and M33 are
star-forming while NGC 205 and M32 are quenched, yielding a quenched
fraction of $40\%$ and conversion fraction of roughly $25\%$. Here, we
define objects in the Local Group and surrounding field population as
quenched according to their observed atomic gas fractions as reported
by \citet{mcconnachie12}, such that objects with ${\rm M}_{\rm
  HI}/\mstar < 0.1$ are classified as quenched. For the field
population surrounding the Local Group, we find that nearly all
galaxies are star-forming, where the rare exceptions (e.g.~Cetus and
Tucana) are likely systems that previously interacted with either the
Milky Way or M31 (i.e.~``backsplash" or ``super-virial" galaxies,
\citealt{Mamon04}, \citealt{Wetzel14}, Garrison-Kimmel et al.~in
prep). This broadly agrees with our observations of low-mass isolated
systems as well as the results of \citet{geha12}.

While at stellar masses of $>10^{8}~\msun$ the satellite population in
the Local Group shows broad agreement with the observed properties of
satellites around our sample of Milky Way-like systems, at lower
stellar masses Local Group satellites exhibit a
dramatically greater quenched fraction and quenching efficiency (see
Fig.~\ref{fig:LG_sats}). With one exception, IC10, every satellite
galaxy of both the Milky Way and M31 below $10^{8}~\msun$ is
quenched. This dramatic increase in satellite quenching efficiency at
low stellar masses ($< 10^{8}~\msun$) points towards a potential
critical mass scale for satellite quenching, such that environmental
quenching is highly efficient for very low-mass satellites. 

Observations of systems in the local Universe comparable to the Local
Group generally confirm the increased satellite quenching efficiency
at low masses. For example, the low-mass ($\sim {\rm a~few} \times
10^{6}~\mstar$) satellites of M81 are universally quenched, with
observed star formation rates of $< 10^{-5}~\msun~{\rm yr}^{-1}$
\citep{kaisin13}. In addition, recent observations of the NGC~4258
group find an overall satellite quenched fraction of $\sim 50\%$ at
$<10^{8}~\msun$, including some objects with blue rest-frame colors
likely to be satellites at stellar masses equal to Fornax and Leo I
\citep{spencer14}. Such observations, however, must be interpreted
under the caveat that there is significantly greater uncertainty in
determining which objects are satellites and which objects belong to
the field in systems beyond the Local Group, where line-of-sight
distances are more poorly constrained. Given that the field is highly
dominated by star-forming systems at these masses, sample
contamination will strongly bias results to systematically lower
quenched fractions.

Altogether, the observational results from our work and others, which
constrain the relative impact of self-quenching and
environmental quenching at low masses, present a challenge to models
of galaxy evolution. Previous studies at higher stellar mass
($>10^{9.5}~\msun$) find that modern semi-analytic models overpredict
the fraction of quenched satellites \citep{weinmann10, wang12}. In
particular, \citet{kimm09} find that while semi-analytic models
overestimate the number density of quenched satellites at high masses,
the same models are generally able to reproduce the observed quenched
fraction in the field, suggesting that conflicts with observations are
largely driven by overprediction of the the impact of environmental
quenching (i.e.~overestimation of the satellite quenching efficiency)
at $>10^{9.5}~\msun$.

Applying our sample selection criteria to the \cite{guo13}
semi-analytic model yields a satellite quenched fraction significantly
elevated with respect to observations at $<
10^{9.5}~\msun$.\footnote{In the SAM, we consider an object
  ``quenched" if it has a SSFR less than $10^{-11}~\rm yr^{-1}$.}
However, the increased number of quenched satellites in the model is
largely driven by a corresponding overprediction of quenched field
systems. At stellar masses $>10^{9.5}~\msun$, where our data overlap
with that of \citet{kimm09}, we similarly find that the model
reproduces the field quenched fraction, while at lower masses the
\citet{guo13} SAM yields an excessive number of quenched
centrals. Thus, while the model does overpredict the efficiency of
satellite quenching (i.e.~${\rm f}_{\rm convert}$) at low masses, the
effect is subdominant. Below $10^{9}~\msun$, the observed conversion
fractions agree with those in the semi-analytic model to within
$50\%$, while the model overpredicts the quenched fraction of field
objects by roughly a factor of $5$. Combined with the results of
\citet{kimm09}, we find that modern models primarily fail to
accurately describe the physics of feedback (i.e.~self-quenching)
within low-mass galaxies and the environmental quenching of high-mass
galaxies.

\section{Conclusions}

In this work, we study the quenching of satellite galaxies in Milky
Way-like systems, primarily focusing on the dependence of quenching
efficiency on both satellite and host mass. Our principal results are
as follows:

\begin{itemize}

\item{The efficiency of satellite quenching is largely independent of
    satellite mass over roughly three orders of magnitude in stellar
    mass, $10^{8}~\msun < \mstar < 10^{11}~\msun$. Comparison to the
    Local Group suggests that satellite quenching efficiency may be
    significantly greater at yet lower stellar masses ($<
    10^{8}~\msun$), perhaps indicating a critical mass for satellite
    quenching.}

\item{Satellite quenching efficiency is well correlated with host halo
    mass, such that satellites of more massive halos are more likely
    to be quenched. A model in which satellite quenching only occurs
    in halos above a given critical halo mass is consistent with
    (\emph{i}) the observed increase in satellite quenching in systems
    with two massive satellites (i.e.~more massive halos) relative to
    those with one massive satellite in addition to (\emph{ii}) the
    higher incidence of quenched satellites around passive hosts
    relative to their star-forming counterparts with one massive
    satellite.}

\item{Discrepancies between the observed quenched fractions of
    low-mass ($< 10^{9.5}~\msun$) field and satellite galaxies and the
    predictions of the \citet{guo13} semi-analytic model are primarily
    driven by overly-effective internal processes (i.e.~feedback or
    self-quenching mechanisms) that yield an overabundance of quenched
    field systems in the models. While the SAM overpredicts the
    efficiency of satellite quenching at low masses, the excess number
    of quenched satellites in the model is largely a product of the
    overabundance of quenched field systems. In contrast, at higher
    masses ($> 10^{9.5}~\msun$), SAMs are generally able to reproduce
    the star-forming properties of field (or central) galaxies, while
    they instead fail to accurately model the environmental quenching
    mechanisms, thereby overpredicting the number of quenched
    satellites at $> 10^{9.5}~\msun$.}

\end{itemize}

\section*{Acknowledgments} 

We thank Marla Geha, Andrew Wetzel and their collaborators for kindly
suppling their observational results in tabular form, as well as for
productive discussions. CW and JSB recognize support from NSF grants
AST-1009973 and AST-1009999.

Funding for the SDSS and SDSS-II has been provided by the Alfred
P. Sloan Foundation, the Participating Institutions, the National
Science Foundation, the U.S. Department of Energy, the National
Aeronautics and Space Administration, the Japanese Monbukagakusho, the
Max Planck Society, and the Higher Education Funding Council for
England. The SDSS Web Site is http://www.sdss.org/.

The SDSS is managed by the Astrophysical Research Consortium for the
Participating Institutions. The Participating Institutions are the
American Museum of Natural History, Astrophysical Institute Potsdam,
University of Basel, University of Cambridge, Case Western Reserve
University, University of Chicago, Drexel University, Fermilab, the
Institute for Advanced Study, the Japan Participation Group, Johns
Hopkins University, the Joint Institute for Nuclear Astrophysics, the
Kavli Institute for Particle Astrophysics and Cosmology, the Korean
Scientist Group, the Chinese Academy of Sciences (LAMOST), Los Alamos
National Laboratory, the Max-Planck-Institute for Astronomy (MPIA),
the Max-Planck-Institute for Astrophysics (MPA), New Mexico State
University, Ohio State University, University of Pittsburgh,
University of Portsmouth, Princeton University, the United States
Naval Observatory, and the University of Washington.

Support for this work was provided by NASA through Hubble Fellowship
grant 51316.01 awarded by the Space Telescope Science Institute,
which is operated by the Asso- ciation of Universities for Research in
Astronomy, Inc., for NASA, under contract NAS 5-26555.

\label{lastpage}
\end{document}